\documentclass[twocolumn,aps,pra,showpacs,superscriptaddress]{revtex4-1}

\usepackage{amsmath}
\usepackage{graphicx}
\usepackage[export]{adjustbox}

\newcommand{\beq}{\begin{equation}}
\newcommand{\eeq}{\end{equation}}
\newcommand{\bea}{\begin{eqnarray}}
\newcommand{\eea}{\end{eqnarray}}

\mathchardef\nss="711B

\def\nss{\mathcal{S}}

\def\be{\begin{eqnarray}}
\def\ee{\end{eqnarray}}

\begin{document}

\title{Fibonacci optical lattices for tunable quantum quasicrystals}
\author{K. Singh}
\email{kevin@physics.ucsb.edu}
\affiliation{Department of Physics, University of California, Santa Barbara, California 93106, USA}
\affiliation{California Institute for Quantum Emulation, Santa Barbara, California 93106, USA}
\author{K. Saha}
\affiliation{California Institute for Quantum Emulation, Santa Barbara, California 93106, USA}
\affiliation{Department of Physics, University of California, Irvine, California 92697, USA}
\author{S. A. Parameswaran}
\affiliation{California Institute for Quantum Emulation, Santa Barbara, California 93106, USA}
\affiliation{Department of Physics, University of California, Irvine, California 92697, USA}
\author{D. M. Weld}
\affiliation{Department of Physics, University of California, Santa Barbara, California 93106, USA}
\affiliation{California Institute for Quantum Emulation, Santa Barbara, California 93106, USA}

\begin{abstract}
We describe a quasiperiodic optical lattice, created by a physical realization of the abstract cut-and-project construction underlying all quasicrystals. The resulting potential is a generalization of the Fibonacci tiling.  Calculation of the energies and wave functions of ultracold atoms loaded into  such a  lattice demonstrate a multifractal energy spectrum, a singular continuous momentum-space structure, and the existence of controllable edge states.  These results open the door to cold atom quantum simulation experiments in tunable or dynamic quasicrystalline potentials, including topological pumping of edge states and phasonic spectroscopy.
\end{abstract}


\maketitle

\section{Introduction}
Quasiperiodicity has a profound impact on electronic structure, playing a role in phenomena ranging from the quantum Hall effect to quasicrystalline ordering.  However, the formation, stability, excitation, and electronic structure of quasiperiodically ordered systems remain incompletely understood.  Open questions include the nature of electronic conductivity or diffusivity, the spectral statistics,  the nature of strongly correlated magnetic states on a quasicrystalline lattice, topological properties of quasicrystals, and even the shape of the electronic wave functions~\cite{janot1997quasicrystals,senechal1996quasicrystals,steinhardtQCbook,hofstadter-fibonacci-butterfly-2007,Thiel-dubois-QCcommentary,QCinAg,AFinQCs,ZilberbergQC,boundaryphenomena2,brouwerpaper,ames-QCs,hofstadter-superlattice-coldatom-proposal}.  

The exquisite controllability of cold atoms makes them a natural choice for experimental investigation of the open questions regarding quasiperiodicity.  Unique features of such experiments would include precisely variable quasiperiodic parameters, tunable interactions, bosonic or fermionic quantum statistics, and the ability to study dynamical phenomena (in modulated or quenched systems, e.g.).  Numerous theoretical proposals have explored the rich physics of quasiperiodically trapped cold atoms~\cite{bermaQPatomoptics,BEC_loc_inQPstructure,bose-glass-penrose-quasicrystal,smitha-QC-article,BECsinQCs,dominik-blochoscs-aperiodiclatts,recentQCproposal,dassarma-topological-harpermodel,hofstadter-superlattice-coldatom-proposal,time-depQHE,demler-quantum-QCs,lewenstein-hofstadtermoth,clarck-bichromatic,generalizedAAH,flickerwezelQCpaper,polishQCpaper}. However, with the exception of some early experiments on non-degenerate atomic gases in 2D quasiperiodic lattices~\cite{atomsin5foldQCopticallattice,grynberg-quasiperiodicOLs}, the dominant application of quasiperiodic or incommensurate potentials in cold atomic physics thus far has been as a convenient proxy for disorder~\cite[e.g.]{inguscio-andersonloc,bloch-mbl}.  The realization of tunable quasicrystalline potentials for cold atoms would open up a broad range of exciting experiments, complementary to those possible with synthesis and characterization of solid-state or photonic quasicrystals.

 \begin{figure}
\begin{center}
 \hspace{.1\columnwidth}\includegraphics[width=0.5\columnwidth]{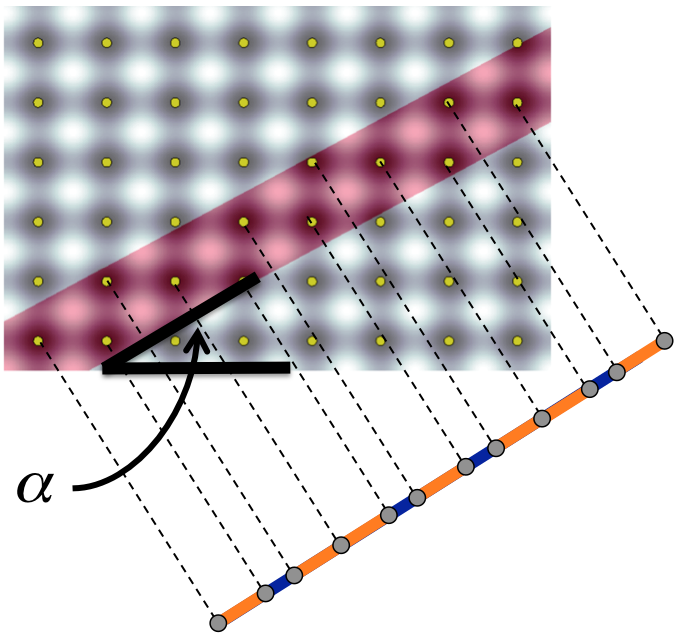} 
\includegraphics[width=\columnwidth]{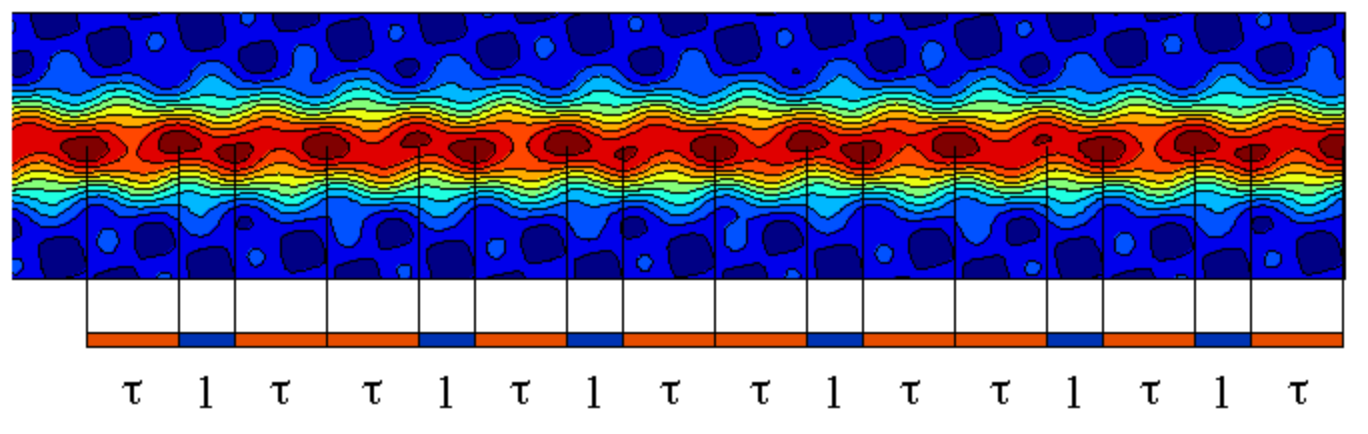}
\end{center}
\vspace{-.2in}
\caption{(Color online) The generalized Fibonacci optical lattice.  \textbf{Top:} Diagram of cut-and-project construction of the generalized Fibonacci lattice. A 2D strip at a particular angle $\alpha$ is projected down to a line.  If $\tan(\alpha)=1/\tau\equiv2/(1+\sqrt{5})$, this results in the Fibonacci tiling itself;  a different irrational slope creates a different 1D quasicrystal. \textbf{Bottom:}  Calculated potential of a Fibonacci optical lattice with parameters discussed in text (red is deeper), showing Fibonacci sequence $\tau 1 \tau \tau 1 \tau 1 \tau \tau 1 \tau \tau 1 \tau 1 \tau$... of lattice spacings.} \label{cutproj}
\end{figure} 
In this paper, we describe and elucidate the properties of a ``generalized Fibonacci'' optical lattice which creates a dynamically tunable family of 1D quasiperiodic structures.  This lattice physically realizes the abstract cut-and-project construction which underlies all quasicrystals.  Every quasiperiodic tiling can  be defined as a projection of a cut through a lattice which is fully periodic but exists in a higher dimensional space~\cite{bohr-cutproject,senechal1996quasicrystals}.   For example, the Fibonacci tiling is a projected 1D cut from a 2D square lattice, and the Penrose tiling can be constructed as a projected 2D cut through a 5D periodic lattice.  In these and all other quasicrystals, the hidden degrees of freedom in the higher-dimensional space can give rise to phenomena such as phasonic excitations and Bragg diffraction with unconventional symmetry~\cite{steinhardtphononsohasons,ZilberbergQC,fleischer-photonicquasicrystals}.   Generalized Fibonacci optical lattices will provide a flexible platform for realization of tunable quantum quasicrystals, and should enable direct experimental  investigation of questions inaccessible to experiments on static, non-tunable, non-interacting systems.  Specific topics of interest include studies of edge states, adiabatic quantum pumping, multifractal energy spectra, phasonic spectroscopy, dynamical signatures of many-body localization, and transport in quasicrystals.

\section{Construction of Generalized Fibonacci Lattices}

The generalized Fibonacci optical lattice is constructed as a direct real-space realization of the cut-and-project procedure, by intersecting an elongated optical trap at a tunable angle with a large-period square lattice, as diagrammed in Fig.~\ref{cutproj}.   The resulting potential is the sum of two simple red-detuned  trap potentials: the large-period square optical lattice, with potential
\begin{align*}
V_{L}(x,y) = -A_{L} \sin^{2}\left(\frac{2\pi x}{\lambda_L}\right) -A_{L} \sin^{2}\left(\frac{2\pi y}{\lambda_L}\right),
\end{align*}
and the elongated cutting beam at an angle $\alpha$ to the $x$ axis, with potential
\begin{align*}
V_{C}(x,y) = -A_{C} \exp\left[-2\left(\frac{-x\sin(\alpha) + y\cos(\alpha)}{\omega_{0}}\right)^{2}\right].
\end{align*}
Here $A_{L}$ is the depth of the square lattice, $\lambda_L/2=a$ is the lattice constant, $A_{C}$ is the trap depth of the cutting beam, $\omega_{0}$ is the beam waist, and we have assumed a Rayleigh range long compared to the trapping region. The natural energy scale is the recoil energy $E_R=h^2/2m\lambda_L^2$. The total potential is then $U(x,y) = V_{L}(x,y)+V_{C}(x,y,\alpha)$.  This potential is shown in Fig.~\ref{cutproj} for $\tan(\alpha)=2/(1+\sqrt{5})$, and $A_C=10 A_L$. In order for the total physical potential to be a good approximation to a true cut-and-project potential, $A_C/A_L$ must be sufficiently large that we can spectrally distinguish states along the cutting beam from transversely-extended states.  To preserve the 1D character of the potential, the waist of the cutting beam should not be large compared to the lattice constant $a$; however, the calculations presented below indicate that the potentials retain many of their interesting quasiperiodic properties even if this condition is violated.  A perturbative treatment elucidating the quasiperiodic character of the effective potential appears in the appendix. This trap construction can be generalized in a straightforward way to 2D quasiperiodic traps, via intersection of a light sheet with a 3D large-period lattice.  This direct experimental realization of the simplest quasiperiodic lattices is also intrinsically tunable: variation of the intersection angle $\alpha$ tunes the properties of the resulting potential, generating different members of this family of quasicrystals, and variation of the offset transverse to the cut beam axis drives phasonic degrees of freedom~\cite{phasonreview}.  

We now briefly discuss the practical optics required to realize such a potential.  If the cutting beam is produced by focusing a gaussian beam of initial diameter $D$ with a lens of focal length $F$, the number of lattice sites in one Rayleigh range of the beam is given approximately by
$$
N_\parallel=\frac{16}{\pi}\frac{\lambda_C}{\lambda_L}\left(\frac{F}{D}\right)^2,
$$
where $\lambda_C$ is the wavelength of the cutting beam and $\lambda_L$ is twice the lattice period.  Even if these traps are produced by the same laser, ${\lambda_C}/{\lambda_L}$ can be varied by using an angled-beam lattice configuration.  The number of lattice sites spanning the cutting beam width is
$$
N_\bot= \frac{8}{\pi}\frac{\lambda_C}{\lambda_L}\left(\frac{F}{D}\right),
$$
so the aspect ratio of the full trap is $N_\parallel/N_\bot=2F/D$.  With typical values of $F$ and $D$, one can then realize a range of generalized Fibonacci traps, with widths ranging from less than a lattice constant to many lattice constants.  The ends of such a trap can be defined for example by tightly-focused blue-detuned light sheets.  The intersection angle $\alpha$ is most easily tuned by rotating the lattice itself; for angled-beam lattices created using a diffractive optical element, this could be straightforwardly achieved with a single rotation stage.  As with ordinary optical lattices, adiabatic loading of cold atoms into a Fibonacci-type lattice would be accomplished starting from the elongated optical trap by a slow turn-on of the lattice potential.

Using this trap geometry, one can construct a continuous family of quasiperiodic tilings of the line by placing the cutting beam at any irrational slope. If the angle of intersection $\alpha$ satisfies the relationship $\tan(\alpha) = 1/\tau$ where $\tau$ is the golden mean $(1+\sqrt{5})/2$, then the resulting potential will approximate the Fibonacci tiling; this is the reason we refer to the family of potentials as ``generalized Fibonacci'' lattices.  This one-dimensional structure tiles the line quasiperiodically, exhibits sharp diffraction peaks, and can also be generated algebraically using the substitution rule $\tau \rightarrow \tau 1$, $1 \rightarrow \tau$, which gives rise to the sequence (1, $\tau$, $\tau 1$, $\tau 1 \tau$, $\tau 1 \tau \tau 1$, $\tau 1 \tau \tau 1 \tau 1 \tau$, $\tau 1 \tau \tau 1 \tau 1 \tau \tau 1 \tau \tau 1$...).  As the bottom panel of Fig.~\ref{cutproj} demonstrates, the energy minima of the Fibonacci optical lattice are spaced according to the Fibonacci tiling.  Because of the inflation symmetry associated with the Fibonacci tiling, if the width of the cut-out strip is reduced, then the resulting one-dimensional projection will simply be an expanded and displaced version of the Fibonacci tiling~\cite{marder}. In the generalized Fibonacci optical lattice, as the width of the Gaussian cutting beam is increased, the potential no longer approximates a one-dimensional projection, but remains quasiperiodic.  The connection to a mathematically exact cut-and-project lattice and the quasiperiodicity of the potential emerge clearly from a perturbative treatment, which also makes plain the connection to closely related systems such as the Aubry-Andr\'e model.  Details of such a perturbative treatment appear in the appendix.  This general optical technique for construction of a family of quasiperiodic lattices and their rational approximants is the first main result of this work.

\begin{figure}[t!]
\begin{center}
\includegraphics[width=\columnwidth]{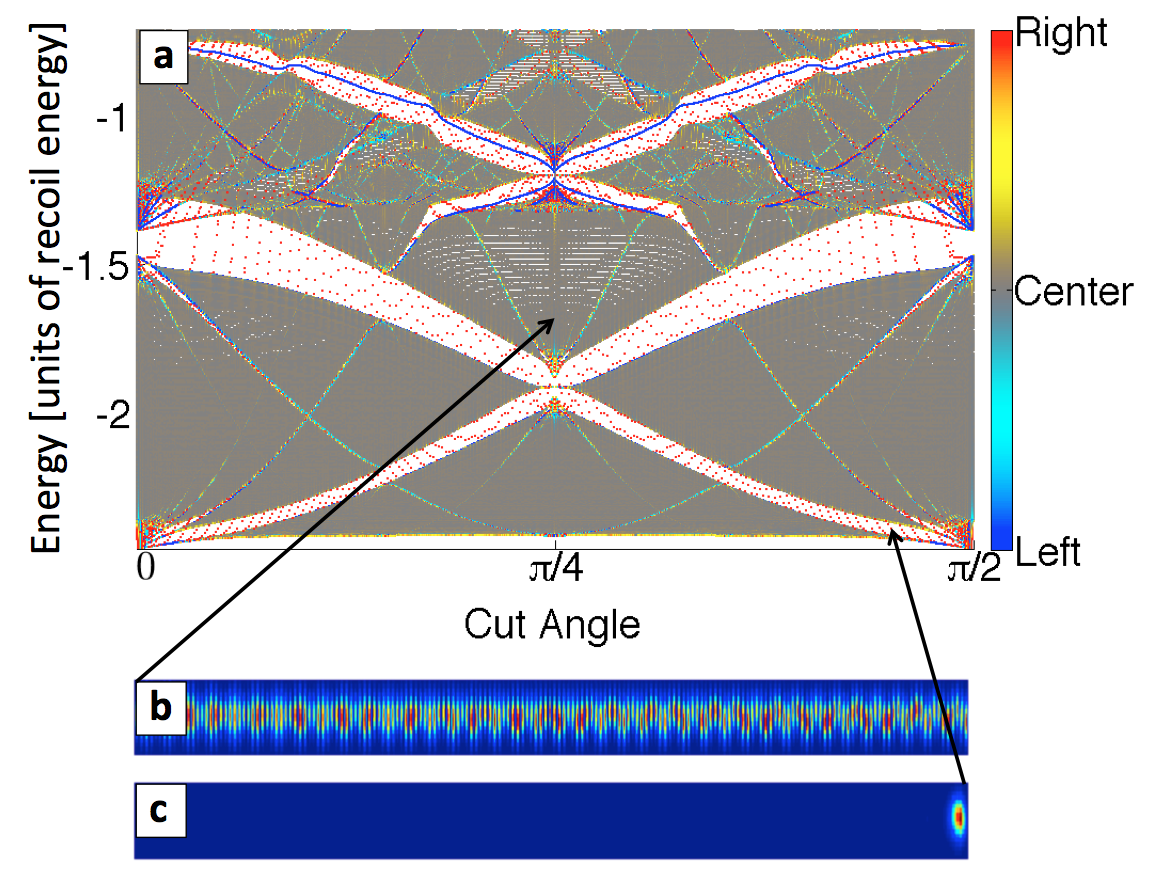}
\end{center}
\caption{(Color online) Energies and wave functions in a tunable Fibonacci-type potential.  \textbf{a:} A portion of the energy spectrum of the generalized Fibonacci optical lattice as a function of cut angle. Here $a=1$, $A_{L} = 5/\pi^2$, $A_{C} = 5A_{L}$, $w_{0} = 1$.  Color of points corresponds to center-of-mass of probability density, to enable identification of edge states.  \textbf{b:}~2D probability density versus position at the indicated cut angle and energy (a typical bulk state).  Red regions correspond to higher probability density. \textbf{c:}~2D probability density versus position at the indicated cut angle and energy (a typical edge state).  Both \textbf{b} and \textbf{c} show a region of 4 by 200 lattice constants.} \label{myenergy}
\end{figure}

\section{wave functions and Energies in Generalized Fibonacci Lattices}

The second main result of this work is the calculation of the energy spectra and wave functions of non-interacting atoms trapped in this family of tunable quasicrystalline potentials.  These calculations demonstrate the utility of generalized Fibonacci optical lattices as a tool for the investigation of quasiperiodic quantum phenomena.  To determine the energy spectrum of the physically-realized trap, we solve the two-dimensional single-particle Schrodinger equation on a mesh with spacing much smaller than a lattice constant.  This approach avoids simplifications inherent in the tight-binding approximation, and makes closer contact with experimentally realizable traps.  We do not use periodic boundary conditions, both for more direct comparison with real experiments and so as to accurately model the existence of edge states at the ends of the quasicrystal.  Energy eigenvalues as a function of cutting beam angle are shown in Fig.~\ref{myenergy}.  The calculated spectrum has a complex multifractal appearance~\cite{multifractalFibOL}.  Notable features include a hierarchy of minigaps which disperse as the angle is varied, a non-accidental resemblance to the Hofstadter butterfly, and the existence of isolated states in the gaps.  We find that the qualitative structure of the energy spectrum remains the same if the waist of the Gaussian beam is increased to several times the size of the lattice constant.  The resemblance to the Hofstadter butterfly is to be expected, given the recent demonstration that the generalized Fibonacci quasicrystal and the Harper model of  high-magnetic-field 2D integer quantum Hall states are topologically equivalent~\cite{equivalence,hofstadter-theoriginal}. The intersection angle $\alpha$ of the Fibonacci quasicrystal plays a role analogous to that of the modulation period of the Harper lattice, or the effective magnetic field in the quantum Hall system. 

\begin{figure}
\begin{center}
\includegraphics[width=.82\columnwidth]{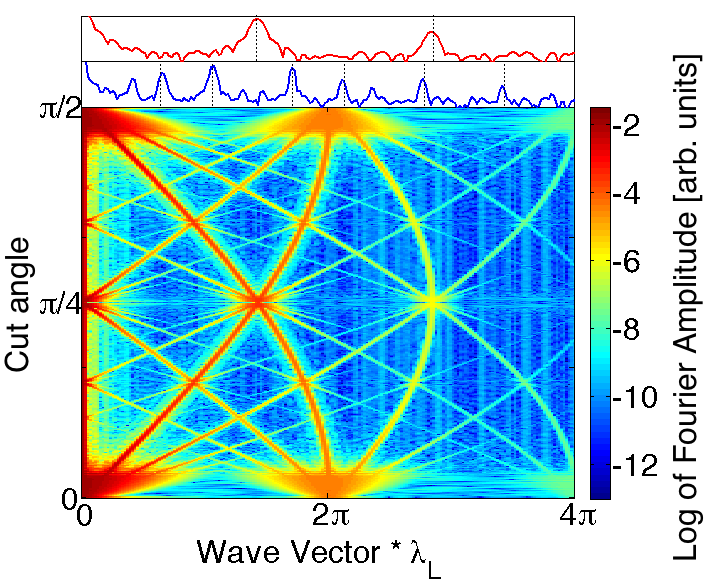}
\end{center}
\caption{(Color online) Tunable quasicrystals in momentum space.  Main lower panel shows Fourier transform of the ground state probability density along the direction of the cutting beam, as a function of cut angle $\alpha$.  Note the logarithmic scale on the colorbar.  Upper panels show log of Fourier transform amplitude versus wave vector at a cut angle of $\pi/4$ (top) and $\tan^{-1}(2/(1+\sqrt{5}))$ (bottom), with identical  axis limits.  Dashed lines show expected wave vectors of Fourier peaks of the potential based on the second-order perturbation theory described in the appendix.  Calculations were performed on a mesh 4 by 200 lattice constants in size.  Faint vertical lines in main plot are finite-size edge effects.} \label{myft}
\end{figure}
The wave functions of atoms in generalized Fibonacci optical lattices also possess unique characteristics.  As the cut angle $\alpha$ is varied, the Fourier transform of the projected spatial probability density of the ground state, plotted in Fig.~\ref{myft}, shows for irrational $\tan(\alpha)$ a rich singular continuous structure characteristic of quasiperiodic structures.  This property is the 1D analogue of the forbidden Bragg diffraction patterns by which 3D quasicrystals were first discovered~\cite{schechtman-originalQCpaper}, and recalls the definition of a quasicrystal as a structure which produces a sharply peaked diffraction pattern but lacks translational symmetry. The singular continuous nature of the spectrum also emerges naturally from a perturbative treatment of the effective potential (see appendix for details).  Rational $\tan(\alpha)=p/q$ produces a crystalline superlattice with a periodicity which depends upon $q$.  The real-space structure of a typical squared wave function in a Fibonacci optical lattice is shown in Fig.~\ref{myenergy}b.  In addition to extended bulk states, isolated states traversing the band gaps are visible in Fig.~\ref{myenergy}a.  Fig.~\ref{myenergy}c shows the squared wave function of a typical gap-traversing state, located inside the lowest energy band gap, and demonstrates that the wave function is localized towards one edge of the lattice. These states, which can occur at both rational and irrational cut slope, are interesting candidates for realizing topological pumping~\cite{ZilberbergQC,hofstadter-superlattice-coldatom-proposal,boundaryphenomena2,brouwerpaper}.

\section{Selected Applications: Topological Pumping and Phason Spectroscopy}

\begin{figure}
\includegraphics[width=\columnwidth,center]{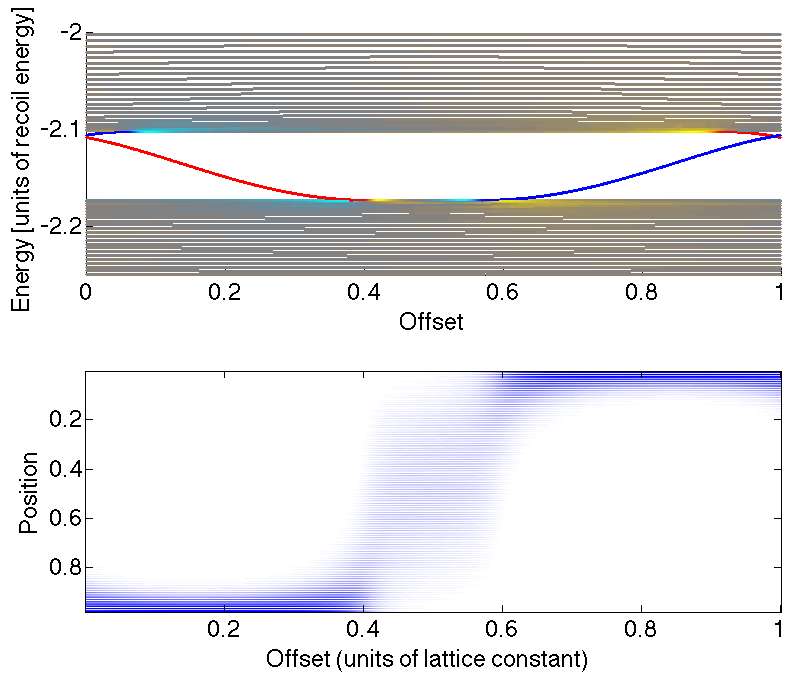}
\caption{(Color online) Edge state topological pumping by phason driving. \textbf{Top:} Energy states near the first minigap versus offset of cutting beam along a lattice vector, at the Fibonacci cut slope.  As the offset between the lattice and the cut beam is varied, left and right edge states cross the gap.  Coloring of points indicates center of mass, with the same mapping as Fig.~\ref{myenergy}.  \textbf{Bottom:} Variation of spatial probability amplitude of an initial edge state as offset is adiabatically varied. Position is normalized to 0 at the left edge and 1 at the right edge.} \label{mydriving}
\end{figure}

Topological pumping is possible because the wave functions and energy spectra of the generalized Fibonacci optical lattice depend in a non-trivial way on the offset of the cutting beam with respect to the lattice.  In the terminology of quasicrystals, this offset is a phasonic degree of freedom.  Just as phonon modes arise from discretely broken real-space translation symmetry, phason modes arise from broken translation symmetry in the higher-dimensional space from which a quasiperiodic lattice is projected~\cite{phasonreview,bakphasons,steinhardtphononsohasons}.  In a Fibonacci-type optical lattice, this corresponds to symmetry under relative translation of the cut beam and the lattice in a direction transverse to the cut beam. A visualization of the effects of continuous adiabatic phasonic driving in the Fibonacci optical lattice is shown in Fig.~\ref{mydriving}.  As the offset of the cutting beam is varied from 0 to 1 lattice constants, an edge state at the right-hand side of the sample with energy in the minigap decreases in energy, merges with the lower band, and later emerges as a left edge state.   These calculations show that adiabatic ramping of the offset can produce a long-range, quantized, oscillatory mass current in a generalized Fibonacci optical lattice.  The effect does not depend on irrationality of the cut slope.  This mass current could be detected, for example, by preferential loading of the edge states in a large-period lattice and direct imaging.  Related effects have recently been observed in photonic waveguide lattices~\cite{ZilberbergQC,fibpumping}, and recent theoretical and experimental work indicates that bulk atomic states can be pumped in a similar way in optical superlattice potentials~\cite{mueller-pumping,thouless-bloch,thouless-takahashi}.  The cold atom context, uniquely,  enables realization of topological pumping in the presence of tunable interactions, and with variable adiabaticity. Such experiments would represent a controllable realization of Thouless pumping of edge states~\cite{thouless-quantizedtransport}, and could provide a powerful tool for dynamical topological control of atomic wave functions.

The availability of the ``hidden dimension'' quite naturally allows us to investigate another aspect of quasicrystalline physics: the role of their soft modes. While of course the optical lattice potentials here are externally imposed and hence do not have true dynamical Goldstone modes, we may simulate the effects of phonons and phasons by suitable manipulations of the lattice and cutting lasers: for instance, shaking the lattice parallel or transverse to the cut corresponds to driving a phonon or a phason, respectively.   Phasons have important but incompletely understood effects on thermal and electronic transport in real quasicrystals~\cite{phasonflips}.  This is of interest not only for fundamental reasons, but also because of potential technological applications of quasicrystals' anomalous electrical and thermal transport characteristics.  The influence of phasons is not understood in large part because of the experimental difficulty of disentangling the effects of domain walls, crystalline impurities, and disorder from those due to phason modes.  A  unique aspect of the generalized Fibonacci optical lattice is that it enables direct oscillatory driving of phason modes.  Measuring the response of the system to driving such modes at variable frequency would constitute a new kind of lattice modulation spectroscopy, in which the modulation occurs in the higher-dimensional space from which the quasiperiodic lattice is projected.  This capability, impossible in other quasiperiodic systems, should allow unprecedentedly specific investigation of phason physics.  

\section{Conclusions}

In conclusion, we have described a tunable quasiperiodic optical lattice,  presented calculations of the properties of quantum gases in such a trap, and shown that this generalized Fibonacci optical lattice will enable experimental realization of topological pumping and phason spectroscopy.  Artificial quasicrystals such as those we propose allow the exploration of arbitrary quasiperiodic geometries, unrestricted by the laws of chemistry.  Creation of a fully tunable  quantum quasicrystal would open the door to a large range of exciting experiments beyond those discussed here, including extensions of these techniques to higher-dimensional quasiperiodic lattices.  The proposed realization of the cut-and-project construction  allows tuning across the dimensional crossover from periodic 2D lattices to  quasiperiodic 1D chains, enabling direct investigation of descendant quasiperiodic phases of well-studied correlated  2D systems.  The unique tools of atomic physics can also enable new types of experiments: Feshbach tuning of the scattering length would allow exploration of the poorly understood role of interactions in quasicrystals~\cite{giamarchiQCinteractions}, and time-varying potentials would enable dynamical experiments impossible in static lattices, such as phason spectroscopy.  Experiments on quasiperiodic optical potentials may ultimately prove complementary to synthesis and characterization of solid and photonic quasicrystals, and could open another conceptual angle of attack on the problem of designing and predicting the properties of these complex materials.  

\textbf{Acknowledgements:}  The authors thank R. Senaratne, Z. Geiger, S. Rajagopal, and K. Fujiwara for helpful discussions. DW and KS acknowledge support from the Office of Naval Research (award  N00014-14-1-0805), the Air Force Office of Scientific Research (award FA9550-12-1-0305), the Army Research Office (award W911NF-14-1-0154), and the Alfred P. Sloan foundation (grant BR2013-110).  All authors are members of the California Institute for Quantum Emulation, supported by a President's Research Catalyst Award (CA-15-327861) from the University of California Office of the President.

\appendix*

\vspace{-2mm}
\section{Effective one-dimensional model for harmonic lattices}
In this appendix, we describe how we may construct systematically more accurate approximations of the effective one-dimensional potential for atomic motion parallel to the cut, order-by-order in perturbation theory. Our notation follows that in the main text.
As a first step, we approximate the cutting beam potential $V_C$ at leading order as a harmonic well,

\be
\begin{split}
\hspace{-.2in}V_C (x,y) & = -A_C e^{-2\left(\frac{r_\perp}{\omega_0}\right)^2} \\
& = -A_C + {2A_C}\left(\frac{r_\perp}{\omega_0}\right)^2 + \mathcal{O}\left(\left(\frac{r_\perp}{\omega_0}\right)^4\right) \\
& \approx -A_C + \frac{1}{2} m\omega_\perp^2 r_\perp^2
\end{split}
\ee
where we have defined $\omega_\perp = \sqrt{\frac{4 A_C}{m \omega_0^2}}$.

We now perform a rotation of the coordinate system, $(x,y) \rightarrow (r_\perp, r_\parallel)$. In the new coordinates, the Hamiltonian is (ignoring an unimportant constant)
\be
H = \frac{p_\perp^2}{2m} +  \frac{p_\parallel^2}{2m} + V_C(r_\perp) + V_L(r_\perp, r_\parallel)
\ee
where
\be
\begin{split}
& V_L(r_\perp, r_\parallel) = \\
&\hspace{.2in}  -A_L^x\sin^2\left(\frac{2\pi}{\lambda_L}\left(r_\parallel\cos\alpha - r_\perp\sin\alpha\right)\right) \\
& \hspace{.2in} -A_L^y\sin^2\left(\frac{2\pi}{\lambda_L}\left(r_\parallel\sin\alpha + r_\perp\cos\alpha\right)\right)\label{eq:rotatedV_L}
\end{split}
\ee
is the 2D periodic lattice potential written in the rotated coordinate system. Note that we have allowed for anisotropic $x$- and $y$- coefficients; this will be relevant to the tunability of the lattice, discussed below.
Assuming that $A_L\ll A_C$, we first solve for the exact eigenstates of the trap potential (the `subbands', to borrow terminology from semiconductor physics); we will then compute the effects of the lattice potential on these eigenstates via perturbation theory. First, we write
\be
H = H_0 + H_1\label{eq:fullHam}
\ee
where 
\be
H_0 = \frac{p_\perp^2}{2m} +  V_C(r_\perp)
\ee
and
\be
H_1 =   \frac{p_\parallel^2}{2m} + V_L(r_\perp, r_\parallel) .
\ee

$H_0$ has exact eigenstates $\psi_n(r_\parallel, r_\perp)$ of the form
\be
\psi_n(r_\parallel, r_\perp) = \chi_{\text{free}}(r_\parallel)\phi_n(r_\perp),\label{eq:separated_form}
\ee
where $\phi_n(r_\perp)$ is an exact eigenstate of the harmonic motion along $r_\perp$:
\be
\left[\frac{p_\perp^2}{2m} + V_C(r_\perp)\right]\phi_n(r_\perp) = \epsilon_n \phi_n(r_\perp), 
\ee
with
\be
\epsilon_n =  \left(n+\frac12\right)\hbar\omega_\perp,
\ee
and $\chi_{\text{free}}$ is any function. Note that we have deliberately chosen to include the kinetic energy along $r_\parallel$ as part of the perturbation $H_1$, as this simplifies the calculations. 

Next, we add in the effects of the $r_\parallel$ dispersion and the lattice potential $V_L$. From (\ref{eq:rotatedV_L}), we see that $V_L$ mixes the transverse and longitudinal motion. It is convenient to account for this by computing corrections to an effective potential for the longitudinal motion $V_{\text{eff}}(r_\parallel)$ order-by-order in perturbation theory. In other words, we restrict the transverse motion to a specified subband (here, we take $n=0$ for specificity, but similar arguments apply, {\it mutatis mutandis}, for any $n$) and compute corrections due to virtual fluctuations to higher subbands order-by-order in perturbation theory. 

For the case of the $n=0$ subband, by a straightforward calculation, this approach yields an effective Hamiltonian for motion along $r_\parallel$,
\be
H_{\text{eff}} = \frac{p_\parallel^2}{2m^*} + V_{\text{eff}}(r_\parallel)
\ee
with $m^*$ the effective mass and
\be
\begin{split}
\hspace{-.4in} & \hspace{-.3in} V_{\text{eff}}(r_\parallel) = \\
&\hspace{-.2in}  V_L^{00}(r_\parallel) - \sum_{n>0} \frac{V_L^{0n}(r_\parallel) V_L^{n0}(r_\parallel)}{n\hbar\omega_\perp} + \mathcal{O}\left(\frac{\|V_L\|^3}{(\hbar\omega_\perp)^2}\right). 
\end{split}
\ee
In the above expression, we have introduced a shorthand for the inter-subband matrix elements of the lattice potential,
\be
V_L^{ln}(r_\parallel) \equiv \int d r_\perp \, \phi^*_l(r_\perp) V_L(r_\parallel, r_\perp) \phi_n(r_\perp).
\ee
It is evident that the perturbative expansion is controlled by the ratio $\frac{\|V_L\|}{(\hbar\omega_\perp)}\approx \frac{A_L}{\hbar \omega_\perp}$ of the inter-subband matrix elements to the subband splitting.  This parameter is straightforwardly tunable in the proposed optical realization of the generalized Fibonacci lattice.  Note that the effective mass will be corrected at higher orders of perturbation theory, and that we may need to be careful about degeneracies introduced at higher orders; a complete analysis of the perturbation theory is beyond the scope of the present work. However, some key qualitative features of the perturbative series, modulo these possible complications, may be extracted simply by studying the form of the perturbation series, as we show in the next section.
\subsection{Fourier analysis of effective potential}
We now turn to an analysis of Fourier components of the effective potential $V_{\text{eff}}$. Consider the $V_{00}$ term. Since there is a single occurrence of $V_L$ at this order, we see that its Fourier transform with respect to $r_\parallel$ will contain only the harmonics present in $V_L$. It is easy to see that at this order, the Fourier transform has Bragg peaks only at $G \in\{ \pm K, \pm K'\}$, where $K = \frac{4\pi}{\lambda_L} \cos\alpha$ and $K' = \frac{4\pi}{\lambda_L}\sin\alpha$ (throughout, we ignore $G = 0$ peaks as they correspond to an unimportant uniform offset of the energy). The corresponding $V_{\text{eff}}$ consists of a pair of harmonics whose minima are respectively at lattice spacings of $\lambda = \frac{\lambda_L}{2\cos\alpha}$  and $\lambda' = \frac{\lambda_L}{2\sin\alpha}$. For $\tan\alpha = 2/(1+\sqrt{5}) =\tau^{-1}$, the ratio of these spacings is indeed the golden ratio, $\lambda'/\lambda = \tau$. It is also straightforward to show that the relative amplitudes of the Bragg peaks is given by
$V_{\pm K} \propto A_L^{x} e^{-\frac14 (K\xi)^2}$, $V_{\pm K'} \propto A_L^{y}e^{-\frac{1}{4}(K'\xi)^2}$, where $
\xi=\sqrt{{\hbar}/{m\omega_{\perp}}}$ is the characteristic `oscillator length' of the trap potential.

\begin{figure}
\includegraphics[width=\columnwidth]{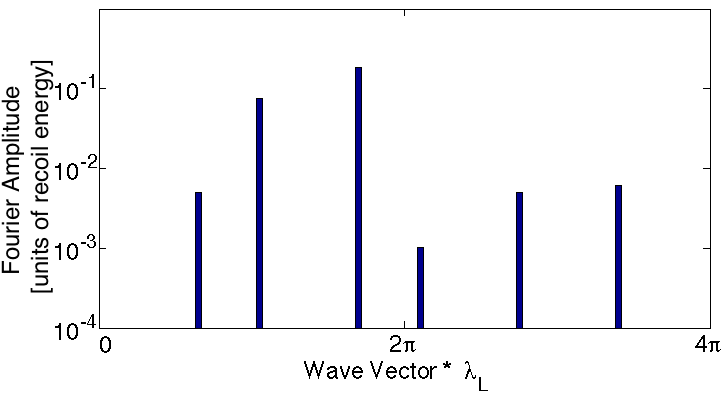}
\caption{(Color online) Fourier spectrum of effective 1D potential. 
The semi-logarithmic plot shows the relative amplitudes of Bragg peaks in units of recoil energy produced by perturbation theory up to second order. We have assumed $A_L^x = A_L^y$.  Note that this is the analytically obtained Fourier transform of the potential itself, rather than the numerically obtained Fourier transform of the atomic density, which is shown in Fig.~3 of the main text.} \label{fourierspectrum}
\end{figure}

At second order, we have two occurrences of $V_L$ in $V_{\text{eff}}$, leading to Bragg scattering at wave vectors $G\in\{ \pm K \pm K', \pm2K, \pm2K'\}$, with the signs all chosen independently. The relative amplitudes  between these peaks are trickier to evaluate analytically, but may be readily computed numerically as needed; however, from the fact that these peaks only emerge at second order, they are accompanied by a factor of $\sim A_L^{x/y}e^{-G^2\xi^2/4}$,
with $x$ or $y$ chosen according to whether we obtain $G$ from the leading peaks by adding $K$ or $K'$, respectively. See Fig.~\ref{fourierspectrum} for a plot of the calculated Fourier spectrum.

We see that we recover a complicated sequence of Bragg peaks as we go to higher orders in perturbation theory. At each order we will find Bragg peaks at higher harmonics, but these will be correspondingly at suppressed amplitude. Generalizing the line of reasoning above, we see that perturbation theory up to order $\mathcal{N}$ yields a set of Bragg peaks
\be
G \in \{ m K + n K' \}
\ee 
where $m,n$ are integers with $|m| + |n| = \mathcal{N}$,
with amplitude  
\be
\hspace{-.3in}f_G\sim \frac{\left(A_L^x\right)^{|m|}\left(A_L^y\right)^{|n|}}{\left(\hbar\omega_\perp\right)^{ |m| + |n| - 1}} \exp\left[ -\frac14 \xi^2\left(m K + n K\right)^2\right].
\ee

The above discussion should make it evident that, at least in principle, the Fourier transform of $V_{\text{eff}}$ is characterized by a singular continuous spectrum, as long as the cut angle is such that $\tan\alpha$ is irrational: in this case, there is no algebraic relation between $K, K'$, and thus the set $\{ m K + n K', \,\,\,\, $m,n$ \in{Z}\}$ has no smallest vector. However, as computed above, and commented on in more detail below, many peaks for large $|m|, |n|$ will have extremely small amplitudes and are for practical purposes absent. Nevertheless, this demonstrates that the generalized Fibonacci optical lattice is strictly more quasicrystalline than the bichromatic potentials created to date, and should be sufficient to explore a variety of relevant physical questions.

\subsection{Comparison to `classical' cut-and-project}
The quick decay of Bragg peak amplitudes at higher orders is a consequence of  the single-harmonic form of the lattice potential and the `soft' projection imposed by the Gaussian beam. Traditional treatments of quasicrystals (see, e.g. \cite{marder}) differ from this in two ways: (i) they discuss a multiple-harmonic lattice, such as the one produced by a periodic delta-function array; and (ii) they incorporate a singular cut (for instance, a step function). As the quasicrystalline potential is the product of these, the Fourier spectrum of the resulting 1D quasicrystal is obtained by convolving the Fourier transform of the lattice, which consists of equal-amplitude peaks at reciprocal lattice vectors, with the slowly decaying Fourier spectrum of a step function (the slow decay is a consequence of the step-edge singularity). The resulting Bragg peaks have an amplitude that decays very slowly (as a power law) with their wave vector. Although this slow decay is known to play an important role in $d=3$ quasicrystals \cite{Kitaev-3dQC}, we note that our approach nevertheless retains the dominant features of quasiperiodicity, namely a set of incommensurate Bragg peaks. Indeed, even purely bichromatic lattices are known to  have a localization-delocalization transition. 

As a final comment, observe that the leading term in our perturbative approach is a direct transcription of cut-and-project: we could have obtained this by convolving the harmonic potential with the Gaussian that describes the atomic density transverse to the cut axis (this is the `tube' that the harmonically confined atoms are restricted to.) However, the higher-order terms (that generate higher Bragg peaks) are in fact non-classical: they emerge due to virtual fluctuations, and have no classical analogue. (Note that a corollary of the previous observation is that the strength of  higher-order Bragg peaks may be enhanced by the inclusion of higher harmonics in $V_L$, as these would then contribute already at leading order.)

\subsection{Relation to Aubry-Andr\'e potential}
A key feature of our set-up is the inherent tunability afforded by the cut-and-project approach. As an example, we sketch a prescription of how to achieve the Aubry-Andr\'e limit in the cut-and-project approach.  To do this, we exploit the ability to impose an anisotropic lattice potential in two dimensions. In the limit $A^y_L=0$, we simply have a stripe modulation in 2D. The effective 1D potential is periodic, with spacing given by $\lambda_L/\cos\alpha$. For a strong Gaussian beam $A_C\gg A_L^x$, we can restrict ourselves to leading order in perturbation theory, so that it suffices to consider the leading harmonic of this 1D potential, with strength $V_0 \propto V_{\pm K} \propto A_L^x$ (here and below, we have ignored factors of $O(1)$ as we are interested in the scaling rather than precise numerical factors). Assuming a relatively deep lattice, we may approximate the projected 1D lattice by a tight-binding chain: in second-quantized form, we have
\be
\hspace{-.3in}H_{\text{eff}} \approx -J\sum_i\left(b^\dagger_i b_{i+1} + b^\dagger_{i+1} b_i\right) - \mu\sum_i b^\dagger_i b_i \label{eq:1DTB}
\ee
with a hopping matrix element (using standard techniques, \cite{zwerger-review}) 
\be
J \approx \frac{4}{\sqrt{\pi}} E_R \left(\frac{V_0}{E_R}\right)^{3/4} \exp\left[-2\left(\frac{V_0}{E_R}\right)^{1/2}\right]
\ee
where  $E_R$ is the effective 1D recoil energy (this is proportional to the recoil energy of the 2D lattice, but as is the case for other quantities, may differ by factors of $O(1)$ due to the projection.)

Next, we impose a weak periodic potential in the $y$ direction, by allowing $A_L^y \neq0$. If we assume that $A_L^x\gg A_L^y\gg (A^x_L)^2/\hbar \omega_\perp$, we may simply consider the additional contribution to the effective potential due to $A^y_L$ as a weak perturbation to the dominant contribution due to  $A^x_L$ (while still ignoring transitions to higher subbands  of the trapping beam). This additional contribution comes from harmonics at $K' = K\tan\alpha$, with strength $V_1 \propto V_{\pm K'} \propto A_L^y$.  Since $A^x_L \gg A^y_L$, this may be incorporated as a weak onsite modulation to the 1D tight-binding Hamiltonian,
\be
\begin{split}
H_{\text{eff}} & \approx \\
& -J\sum_i\left(b^\dagger_i b_{i+1} + b^\dagger_{i+1} b_i\right) \\
& +\sum_i (-\mu + V_1\cos(K' r_i)) b^\dagger_i b_i,  \label{eq:1DTBmod}
\end{split}
\ee
which is equivalent to that of the Aubry-Andr\'e model with a phase offset of zero. We may recover the full Aubry-Andr\'e model by also including an offset phase in the $A_L^y$ term of Eq.(\ref{eq:rotatedV_L}).

%

\end{document}